\documentclass[runningheads]{llncs}
\usepackage{graphicx}
\usepackage{cite}  %
\usepackage{letltxmacro}  %
\usepackage{amssymb}%
\usepackage{amsmath}%
\usepackage{verbatim}
\usepackage{color}%
\usepackage{multirow}
\usepackage[table,xcdraw]{xcolor}%
\begin{document}
\title{Conditional generator and multi-source correlation guided brain tumor segmentation with missing MR modalities}
\titlerunning{Brain tumor segmentation with missing MR modalities}
\author{Tongxue Zhou\inst{1,2,3} \and
St\'ephane Canu\inst{2,3}
\and Pierre Vera\inst{4}
\and Su Ruan\inst{1,3}}
\authorrunning{Z. Tongxue et al.}
\institute{Universit\'e de Rouen Normandie, LITIS - QuantIF, Rouen, France \and
INSA Rouen, LITIS - Apprentissage, Rouen, France \and 
Normandie Univ, INSA Rouen, UNIROUEN, UNIHAVRE, LITIS, France \and
Department of Nuclear Medicine, Henri Becquerel Cancer Center, Rouen, France
}

\maketitle              

\begin{abstract}
Brain tumor is one of the most high‐risk cancers which causes the 5-year survival rate of only about 36\%. Accurate diagnosis of brain tumor is critical for the treatment planning. However, complete data are not always available in clinical scenarios. In this paper, we propose a novel brain tumor segmentation network to deal with the missing data issue. To compensate for missing data, we propose to use a conditional generator to generate the missing modality under the condition of the available modalities. As the multi-modality has a strong correlation in tumor region, we design a correlation constraint network to leverage the multi-source information. On the one hand, the correlation constraint network can help the conditional generator to generate the missing modality which should keep the multi-source correlation with the available modalities. On the other hand, it can guide the segmentation network to learn the correlated feature representations to improve the segmentation performance. The proposed network consists of a conditional generator, a correlation constraint network and a segmentation network. We carried out extensive experiments on BraTS 2018 dataset to evaluate the proposed method. The experimental results demonstrate the importance of the proposed components and the superior performance of the proposed method compared with the state-of-the-art methods.
\end{abstract}

\section{Introduction}
Brain tumor is one of the most fatal cancers in the world. Accurate brain tumor segmentation is of critical importance to improve diagnosis, perform surgery and make treatment planning. MRI is the first choice for brain tumor diagnosis and treatment planning, because it has the superior image contrast in soft tissue. In addition, different MR modalities can provide complementary information to improve the diagnosis results~\cite{menze2014multimodal}.


In recent years, many deep learning based methods are proposed to automatically segment brain tumors based on MRI data \cite{myronenko20183d, akil2020fully, ding2020multi, zhou2020one}. These methods apply the full modalities to do the segmentation. However, it usually happens to have missing data in clinical scenarios. Recently, a lot of researches are conducted to segment the brain tumor with missing data. For example, Havaei et al.~\cite{havaei2016hemis} proposed to first project each available image into a single latent representation space. Then, these single latent representations are merged via computing the mean and variance to achieve the segmentation. Similarly, Lau et al.~\cite{lau2019unified} proposed to map a variable number of input modalities into a unified representation by calculating the mean for the final segmentation. To further enhance the modality-invariance of latent feature, Chartsias et al.~\cite{chartsias2017multimodal} proposed to minimize the L1 or L2 distance of features from different modalities. Since different MRI modalities have different intensity distributions. Therefore, using arithmetic operations or simply encouraging the features from different modalities to be close under L1 or L2 distance, could not guarantee that the network can learn a shared latent representation. To this end, Dorent et al.~\cite{10.1007/978-3-030-32245-8_9} proposed to use multi-modal variational auto-encoders to combine the available modalities to a shared latent representation for brain tumor segmentation. Recently, conditional Generative Adversarial Network (cGAN) \cite{mirza2014conditional, yu20183d} has demonstrated to be a promising approach for image synthesis. However, the training of GAN is highly unstable and difficult to converge. Currently, U-Net has been widely used in the synthesis task due to its contracting and expanding paths in the encoder and decoder. Different from the previous works \cite{emami2020frea, han2017mr} where a native U-Net is used as the generator, we propose a conditional U-Net to generate the missing modality from the available ones. Specifically, we use the missing modality index as the condition. The condition imposed on the generator makes it possible to generate the missing modality in a more relevant and supervised way. It is noted that our task is not to generating a perfect image but to do the segmentation. We focus on learning the effective features for segmentation. Benefiting from the generated modality, we can obtain an additional feature information to improve the segmentation.

In this paper, we propose to combine conditional generator and multi-source correlation to help the brain tumor segmentation in the case of missing data. The preliminary conference versions appeared at MICCAI 2020~\cite{zhou2020brain}. It replaces a missing modality by an available modality. This paper extended the previous work by designing an additional conditional generator to generate the missing modality. The whole network architecture is totally different. The main contributions of our method are: 1) A conditional generator is designed to generate the missing modality with correlation constraint. 2) A correlation constraint network is introduced to extract the latent multi-source correlation between modalities, helping the conditional generator to emphasize correlated tumor features of the missing modality. 3) The brain tumor segmentation network utilizing the multi-source correlation to conditionally generate the missing modality is proposed.

\begin{figure}
\centering
\includegraphics[width=\textwidth]{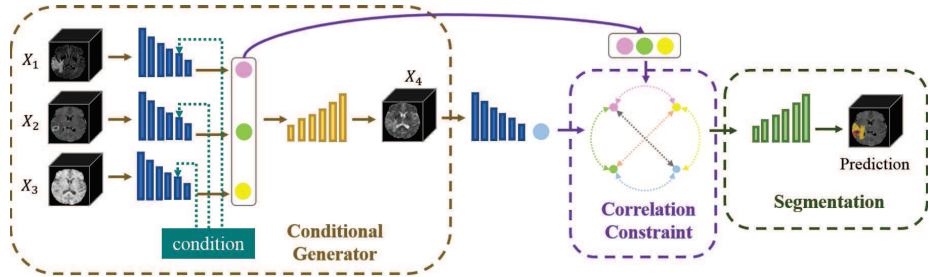}
\caption{The overview of our proposed network, consisting of a conditional generator, a correlation constraint network and a segmentation network, here $X_4$ represents the missing modality.}
\label{fig1}
\end{figure}

\section{Method}
The proposed network consists of three sub-networks, a conditional generator, a correlation constraint network and a segmentation network. It is trained in an end-to-end fashion. The overview of the network is depicted in Fig. \ref{fig1}. First,  the conditional generator takes the available modalities as inputs to generate the missing modality. Here, the condition is the missing modality index, which is inserted to the last second layer of each encoder. Then, an additional encoder is used to extract the individual feature representation from the generated modality. Following that, the correlation constraint network takes the individual feature representations of the full modalities to discover the multi-source correlation between modalities. Finally, a decoder is introduced to do the final segmentation. 

\subsection{Conditionally generating the missing modality} 
\begin{figure}
\centering
\includegraphics[width=0.8\textwidth]{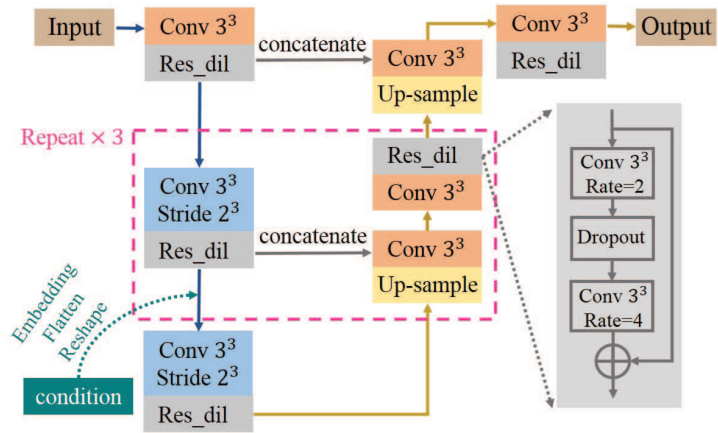}
\caption{The architecture of the proposed conditional generator. We take one encoder and the decoder as an example. The left series of blocks connected by the blue arrows represent the encoder, and right ones connected by the yellow arrows represent the decoder.}
\label{fig2}
\end{figure}

The proposed conditional generator is a multi-encoder based U-Net. The architecture is depicted in Fig.~\ref{fig2}. Specifically, each encoder consists of a convolutional layer and a res\_dil block in each level. The res\_dil block is inspired by the residual block and dilated convolution. It is used to increase the receptive field so as to learn more semantic features. In the last second layer of the encoder, the condition (missing modality index) is added as an additional input layer. We refer to the condition coding method in cGAN~\cite{mirza2014conditional}. Here, we propose to encode the missing modality as an index ($0, 1, 2, 3$ corresponding to T2, T1c, FLAIR and T1 respectively.) using an embedding layer. Then, we flatten and reshape it to the same size of the feature maps in the encoder. Finally, it is concatenated with the feature maps. The condition can constrain the generator to generate the corresponding missing modality in a more targeted way. The decoder begins with a up-sampling layer and a convolution layer. Then the up-sampled features are concatenated with the features from the corresponding level of the encoders. Following the concatenation, a convolution layer is first applied to adjust the number of features, and then the res\_dil block is used to enlarge the receptive field.

It is noted that, the conditional generator and the following segmentation network share the same encoders for the available modalities. The advantages are three-fold: (1) The sharing operation can simplify the network architecture and reduce the training parameters. (2) The conditional generator can learn the tumor related feature information from the segmentation network, making the interested regions enhancing in the generated modality. (3) The segmentation network joins the generated modality with other available modalities to improve the segmentation performance.

\subsection{Discovering the multi-source correlation}
\begin{figure}
\centering
\includegraphics[width=\textwidth]{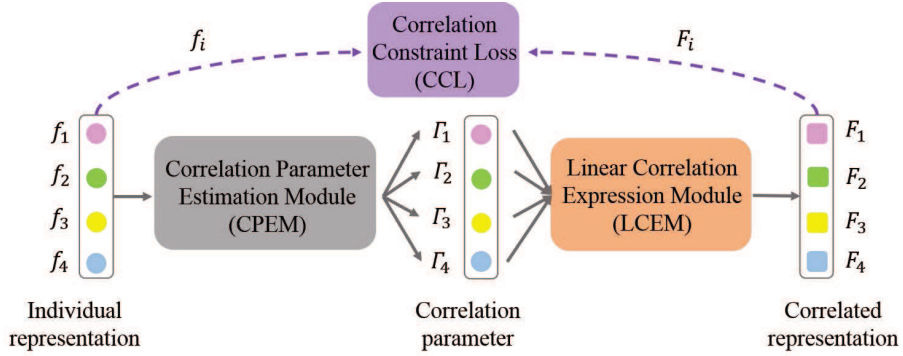}
\caption{The architecture of the proposed correlation constraint network, consisting of CPEM, LCEM and CCL.}
\label{fig3}
\end{figure}

Since the same region of interest (tumor region) can be observed in different modalities for each patient. Therefore, there is a strong correlation in intensity distribution between each pair of modalities~\cite{zhou2020brain}. To this end, we introduce a Correlation Constraint (CC) network to discover the multi-source correlation between modalities. The detailed network architecture is presented in Fig.~\ref{fig3}. The proposed CC network consists of three components: Correlation Parameter Estimation Module (CPEM), Linear Correlation Expression Module (LCEM) and Correlation Constraint Loss (CCL). After generating the missing modality $X_4$, a new set of the complete modalities are obtained. Each input modality $\{X_i\}$, where $i=\{1,2,3,4\}$, is input to the independent encoder to learn the individual feature representation $f_i(X_i|\theta_i)$, where $\theta_i$ denotes the parameters used in $ith$ encoder, such as the number of filters, the kernel size of filter and the rate of dropout. It is noted that only the generated modality needs to train an additional encoder to get the independent feature representation, the other encoders are directly taken from the generator. Then, the CPEM is used to map the individual feature representation $f_i(X_i|\theta_i)$ to a set of correlation parameters $\Gamma_i =\{\alpha_i, \beta_i, \gamma_i, \delta_i\}$. CPEM is a network with two fully connected networks. Finally, the LCEM is employed to produce the correlated feature representation $F_i(X_i|\theta_i)$ for each modality (Equation~\ref{eq1}). The correlation expression in our work is linear. However, the proposed LCEM can be generally integrated to any multi-source correlation problem, and the correlated expression will depend on the application. Finally, a Kullback–Leibler (KL) divergence based CCL (Equation~\ref{eq2}) is introduced. It can constrain the distributions between the estimated correlated feature representation and the original feature representation to be as close as possible. 

In conclusion, on the one hand, the CC network can help the conditional generator to generate the missing modality which should keep the multi-source correlation with the available modalities. On the other hand, it can guide the segmentation network to learn the correlated feature representations to improve the segmentation performance.

\begin{equation}
    F_i(X_i|\theta_i) = \alpha_i \odot f_j(X_{j}|\theta_{j})+\beta_i \odot f_k(X_{k}|\theta_{k})+\gamma_i \odot f_l(X_{l}|\theta_{l})+\delta_i)
    , (i \neq j \neq k \neq l)
\label{eq1}
\end{equation}

\noindent where $X$ is the input modality, $i$, $j$, $k$ and $l$ are the indexes of the modality, $\theta$ is the network parameters, $f$ is the individual feature representation, $F$ is the correlated feature representation, $\alpha$, $\beta$, $\gamma$ and $\delta$ are the correlation parameters.

\begin{equation}
 L_{cc} = \sum_{i=1}^M {P_i} log\frac{P_i}{Q_i}
\label{eq2}
\end{equation}
\noindent where $M$ is the number of modality, $P_i$ and $Q_i$ are the original feature representation distributions and correlated feature representation distributions of modality $i$, respectively.

\begin{table}
\caption{Comparison of segmentation in terms of Dice Similarity Coefficient (DSC) and Hausdorff Distance (HD) on BraTS 2018 dataset. AVG denotes the average results on the three regions, bold results denote the best scores.}
\label{tab1}
\begin{center}
\resizebox{\textwidth}{!}{%
\begin{tabular}{c|c|cccc|cccc|cccc|cccc}
\hline
\multicolumn{2}{c|}{} & \multicolumn{4}{c|}{Missing FLAIR} & \multicolumn{4}{c|}{Missing T1} & \multicolumn{4}{c|}{Missing T1c} & \multicolumn{4}{c}{Missing T2} \\ \cline{3-18} 
\multicolumn{2}{c|}{\multirow{-2}{*}{Methods}} & WT & TC & ET & AVG & WT & TC & ET & AVG & WT & TC & ET & AVG & WT & TC & ET & AVG \\ \hline
 & \cellcolor[HTML]{EFEFEF}DSC & \cellcolor[HTML]{EFEFEF}55.1 & \cellcolor[HTML]{EFEFEF}53.6 & \cellcolor[HTML]{EFEFEF}67.5 & \cellcolor[HTML]{EFEFEF}58.7 & \cellcolor[HTML]{EFEFEF}84.2 & \cellcolor[HTML]{EFEFEF}77.6 & \cellcolor[HTML]{EFEFEF}69.7 & \cellcolor[HTML]{EFEFEF}77.2 & \cellcolor[HTML]{EFEFEF}85.7 & \cellcolor[HTML]{EFEFEF}45.8 & \cellcolor[HTML]{EFEFEF}2.1 & \cellcolor[HTML]{EFEFEF}44.5 & \cellcolor[HTML]{EFEFEF}81.8 & \cellcolor[HTML]{EFEFEF}69.8 & \cellcolor[HTML]{EFEFEF}72.6 & \cellcolor[HTML]{EFEFEF}74.7 \\ \cline{2-18} 
\multirow{-2}{*}{Replace} & HD & 37.9 & 52.8 & 14.1 & 34.9 & 7.0 & 6.9 & 4.9 & 6.3 & 6.4 & 16.0 & 24.8 & 15.7 & 9.0 & 11.8 & 3.7 & 8.2 \\ \hline
 & \cellcolor[HTML]{EFEFEF}DSC & \cellcolor[HTML]{EFEFEF}75.5 & \cellcolor[HTML]{EFEFEF}76.5 & \cellcolor[HTML]{EFEFEF}71.9 & \cellcolor[HTML]{EFEFEF}74.6 & \cellcolor[HTML]{EFEFEF}84.6 & \cellcolor[HTML]{EFEFEF}83.7 & \cellcolor[HTML]{EFEFEF}76.4 & \cellcolor[HTML]{EFEFEF}81.6 & \cellcolor[HTML]{EFEFEF}84.8 & \cellcolor[HTML]{EFEFEF}62.1 & \cellcolor[HTML]{EFEFEF}\textbf{40.1} & \cellcolor[HTML]{EFEFEF}62.3 & \cellcolor[HTML]{EFEFEF}84.9 & \cellcolor[HTML]{EFEFEF}84.2 & \cellcolor[HTML]{EFEFEF}76.8 & \cellcolor[HTML]{EFEFEF}82.0 \\ \cline{2-18} 
\multirow{-2}{*}{Direct} & HD & 10.4 & 9.0 & 6.2 & 8.5 & 7.6 & 6.9 & 4.4 & 6.3 & 6.8 & 12.4 & \textbf{10.5} & 9.9 & 6.4 & 6.4 & 4.0 & 5.6 \\ \hline
 & \cellcolor[HTML]{EFEFEF}DSC & \cellcolor[HTML]{EFEFEF}79.9 & \cellcolor[HTML]{EFEFEF}78.8 & \cellcolor[HTML]{EFEFEF}72.2 & \cellcolor[HTML]{EFEFEF}77.0 & \cellcolor[HTML]{EFEFEF}85.1 & \cellcolor[HTML]{EFEFEF}83.4 & \cellcolor[HTML]{EFEFEF}77.1 & \cellcolor[HTML]{EFEFEF}81.9 & \cellcolor[HTML]{EFEFEF}85.7 & \cellcolor[HTML]{EFEFEF}62.8 & \cellcolor[HTML]{EFEFEF}\textbf{40.1} & \cellcolor[HTML]{EFEFEF}62.9 & \cellcolor[HTML]{EFEFEF}85.4 & \cellcolor[HTML]{EFEFEF}85.1 & \cellcolor[HTML]{EFEFEF}77.3 & \cellcolor[HTML]{EFEFEF}82.6 \\ \cline{2-18} 
\multirow{-2}{*}{Direct+CC} & HD & 7.9 & 10.3 & 8.1 & 8.8 & 6.5 & 5.3 & 3.0 & 4.9 & 6.7 & \textbf{11.8} & 10.9 & 9.8 & 6.8 & 6.0 & 3.8 & 5.5 \\ \hline
 & \cellcolor[HTML]{EFEFEF}DSC & \cellcolor[HTML]{EFEFEF}\textbf{83.5} & \cellcolor[HTML]{EFEFEF}\textbf{85.0} & \cellcolor[HTML]{EFEFEF}\textbf{76.5} & \cellcolor[HTML]{EFEFEF}\textbf{81.7} & \cellcolor[HTML]{EFEFEF}\textbf{87.0} & \cellcolor[HTML]{EFEFEF}\textbf{86.2} & \cellcolor[HTML]{EFEFEF}\textbf{77.7} & \cellcolor[HTML]{EFEFEF}\textbf{83.6} & \cellcolor[HTML]{EFEFEF}\textbf{86.5} & \cellcolor[HTML]{EFEFEF}\textbf{64.0} & \cellcolor[HTML]{EFEFEF}39.2 & \cellcolor[HTML]{EFEFEF}\textbf{63.2} & \cellcolor[HTML]{EFEFEF}\textbf{87.1} & \cellcolor[HTML]{EFEFEF}\textbf{86.8} & \cellcolor[HTML]{EFEFEF}\textbf{78.2} & \cellcolor[HTML]{EFEFEF}\textbf{84.0} \\ \cline{2-18} 
\multirow{-2}{*}{\begin{tabular}[c]{@{}c@{}}Direct+CC+CG\\ (Ours)\end{tabular}} & HD & \textbf{6.4} & \textbf{4.8} & \textbf{3.5} & \textbf{4.9} & \textbf{4.5} & \textbf{4.2} & \textbf{2.9} & \textbf{3.9} & \textbf{5.3} & 12.0 & 11.1 & \textbf{9.5} & \textbf{4.5} & \textbf{3.5} & \textbf{2.3} & \textbf{3.4} \\ \hline
\end{tabular}%
}
\end{center}
\end{table}
\subsection{Brain tumor segmentation}
\label{sec2.3}
We applied our previous work~\cite{zhou2020multi}, a multi-encoder based U-Net, to do the brain tumor segmentation. The multiple encoders can help to extract the independent feature representations for each modality. In addition, an attention mechanism based fusion block is introduced to learn feature representations along spatial-wise and modality-wise. Furthermore, the deep supervision is applied in the decoder by integrating the segmentation results from different levels to form the final network output.

\section{Experiments}
\subsubsection{Dataset and pre-processing.}
BraTS 2018 dataset is used to evaluate our method. It contains 285 cases, each case has four image modalities including T1, T1c, T2 and FLAIR. There are three segmentation classes: whole tumor (WT), tumor core (TC) and enhancing tumor region (ET). The images are cropped and resized from $155 \times 240 \times240$ to $128 \times 128 \times128$. The N4ITK~\cite{avants2009advanced} method is used to correct the distortion of MRI data, and intensity normalization is applied to normalize each modality to a zero-mean, unit-variance space. 

\begin{table}
\caption{Comparison of different methods in terms of Dice Similarity Coefficient (DSC) on BraTS 2018 dataset. AVG denotes the average results on the three regions, bold results denote the best scores.}
\label{tab2}
\begin{center}
\resizebox{\textwidth}{!}{%
\begin{tabular}{c|cccccccccccccccc}
\hline
 & \multicolumn{16}{c}{DSC (\%)} \\ \cline{2-17} 
Methods & \multicolumn{4}{c|}{Missing FLAIR} & \multicolumn{4}{c|}{Missing T1} & \multicolumn{4}{c|}{Missing T1c} & \multicolumn{4}{c}{Missing T2} \\ \cline{2-17} 
 & WT & TC & ET & \multicolumn{1}{c|}{AVG} & WT & TC & ET & \multicolumn{1}{c|}{AVG} & WT & TC & ET & \multicolumn{1}{c|}{AVG} & WT & TC & ET & AVG \\ \hline
HeMIS~\cite{havaei2016hemis} & 44.2 & 46.6 & 55.1 & \multicolumn{1}{c|}{48.6} & 75.6 & 54.9 & 60.5 & \multicolumn{1}{c|}{63.7} & 75.2 & 18.7 & 1.0 & \multicolumn{1}{c|}{31.6} & 70.2 & 48.8 & 60.9 & 60.0 \\ \hline
U-HeMIS~\cite{havaei2016hemis} & 82.1 & 70.7 & 69.7 & \multicolumn{1}{c|}{74.2} & 87 & 72.2 & 69.7 & \multicolumn{1}{c|}{76.3} & 87.0 & 61.0 & 33.4 & \multicolumn{1}{c|}{60.5} & 85.1 & 70.7 & 69.9 & 75.2 \\ \hline
URN~\cite{lau2019unified} & 81.1 & 69.5 & 68.5 & \multicolumn{1}{c|}{73.0} & 86.5 & 72.2 & 69.8 & \multicolumn{1}{c|}{76.2} & 86.1 & 52.5 & 25.8 & \multicolumn{1}{c|}{54.8} & 85.6 & 72.0 & 71.0 & 76.2 \\ \hline
HVED~\cite{10.1007/978-3-030-32245-8_9} & 83.3 & 75.3 & 71.1 & \multicolumn{1}{c|}{76.6} & \textbf{88.6} & 75.6 & 71.2 & \multicolumn{1}{c|}{78.5} & \textbf{88.0} & 61.5 & 34.1 & \multicolumn{1}{c|}{61.2} & 86.2 & 74.2 & 71.1 & 77.2 \\ \hline
Ours&\textbf{83.5}&\textbf{85.0} & \textbf{76.5}&\multicolumn{1}{c|}{\textbf{81.7}} &87.0&\textbf{86.2}&\textbf{77.7}& \multicolumn{1}{c|}{\textbf{83.6}} & 86.5&\textbf{64.0}&\textbf{39.2}& \multicolumn{1}{c|}{\textbf{63.2}}&\textbf{87.1}&\textbf{86.6}&\textbf{78.2} &\textbf{84.0} \\ \hline
\end{tabular}%
}
\end{center}
\end{table}

\subsubsection{Implementation details.}
The proposed network is implemented by Keras with a single Nvidia Tesla V100 (32G). The model is trained using Nadam optimizer, the initial learning rate is 0.0005, it will reduce with a factor 0.5 with patience of 10 epochs. To avoid over-fitting, early stopping is used if the validation loss is not improved over 20 epochs. We randomly split the dataset into 80\% training and 20\% testing. All the results are obtained by online evaluation platform\footnote{https://ipp.cbica.upenn.edu/}. The overall loss function used in our network is defined in Equation~\ref{eq3}. Dice loss is used as the segmentation loss, L1-Norm as the generation loss, and KL as the correlation loss.
\begin{equation}
    \ L_{total}= L_{dice} + 0.1\times L_{1} + 0.1\times L_{cc}
\label{eq3}
\end{equation}

\subsubsection{Ablation studies.}
To demonstrate that the effectiveness of our proposed method. First, we compare our method with two related methods. (1) Replace: The idea is from our previous work~\cite{zhou2020miccai}, which trains the network on full modalities but uses the most correlated available modality to replace the missing one during test. (2) Direct: It directly uses the available modalities to do the segmentation. Then, we gradually add the proposed components (CC: correlation constraint and CG: conditional generator) to the 'Direct' method to prove the effectiveness of the proposed components. Dice Similarity Coefficient (DSC) and Hausdorff Distance (HD) are used as the evaluation metrics. From Table~\ref{tab1}, we can observe that the 'Replace' method can't have a satisfying result, especially when FLAIR or T1c is missing. We explain that simply replacing the missing modality with the available one can't compensate the real missing modality. The 'Direct' method can outperform the 'Replace' method, but the results are still unsatisfying. However, when the multi-source correlation is taken into account (Direct + CC), the segmentation results are much improved, especially when FLAIR is missing. In addition, when the conditional generator is used to compensate the missing modality, we can observe a significant improvement of 9.5\% in terms of average DSC compared with the baseline ('Direct') when FLAIR is missing. The similar comparison results can be observed with regard to Hausdorff Distance. In conclusion, the comparison results demonstrate the importance of the proposed components. 
Furthermore, we compare our method with the state-of-the-art methods, which have been mentioned in the introduction. The comparison results are illustrated in Table~\ref{tab2}. Since the method HeMIS \cite{havaei2016hemis} didn't publish the available code, the reported results on HeMIS and U-HeMIS are taken from the work in \cite{10.1007/978-3-030-32245-8_9}. for all the tumor regions, our method achieves the best results in most of the cases. Compared to the current state-of-the-art method~\cite{10.1007/978-3-030-32245-8_9}, our method is superior than it with a large margin. We explain that the proposed conditional generator can compensate the missing modality under the correlation constraint between modalities. 

\subsubsection{Qualitative Analysis.}
\begin{figure}
\centering
\includegraphics[width=1.0\textwidth]{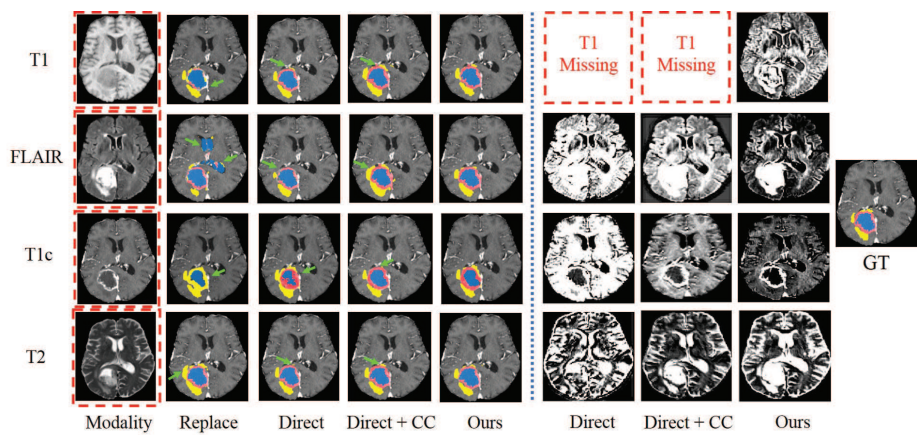}
\caption{Visualization of the segmentation results and feature maps. Left: Each row presents the segmentation results from different methods in the case of missing one modality. For example, the first row presents the segmentation results in the case of missing T1 modality. Right: Each column presents the feature maps obtained by the three methods corresponding to T1, Flair, T1c and T2, respectively, shown from top to bottom. The green arrow highlights the segmentation differences among the different methods. Non-enhancing tumor and necrotic tumor are shown in blue, edema in yellow and enhancing tumor in red.}
\label{fig4}
\end{figure}

To further demonstrate the segmentation performance of our method, we visualize an example from BraTS 2018 dataset in Fig.~\ref{fig4} (the left part). We can observe that the 'Replace' method produces many false predictions, especially when FLAIR and T1c are missing. 'Direct' method can achieve the better results than 'Replace'. Furthermore, the results can be refined progressively using the proposed components with regard to correlation constraint network and the condition generator. In addition, we visualize the feature maps of different methods in Fig.~\ref{fig4} (the right part). Here, we take 'missing T1 modality' as an example. It can be observed that, in the second column, the target tumor regions are really obvious. We explain that the correlation constraint network can learn the related segmentation features. In addition, with the help of the conditional generator, an additional feature information is obtained. It benefits the network to achieve better segmentation result.

\section{Conclusion}
We propose a novel brain tumor segmentation network to deal with the missing data issue. Based on the multi-source correlation between modalities, we first apply the conditional generator to generate the missing modality. Then, a multi-encoder based segmentation network is applied on the new complete modality set to do the segmentation. Experimental results demonstrate the effectiveness of our method. In future work, we are going to improve the generator and extend it to cover missing any number of modalities.

\vfill
\pagebreak
\bibliographystyle{splncs04}
\bibliography{strings}

\end{document}